\documentclass[twocolumn,showpacs,amsmath,amssymb,pra,superscriptaddress,floatfix]{revtex4}

\usepackage{amssymb}
\usepackage{graphicx}% Include figure files
\usepackage{dcolumn}% Align table columns on decimal point
\usepackage{bm}% bold math
\usepackage{psfrag}
\begin{document}

\title{Collective Rydberg excitations of an atomic gas confined in a ring lattice}

\author{B. Olmos}
\email{bolmos@ugr.es}
\affiliation{Instituto 'Carlos I' de F\'{\i}sica Te\'orica y Computacional
and Departamento de F\'{\i}sica At\'omica, Molecular y Nuclear,
Universidad de Granada, E-18071 Granada, Spain}
\author{R. Gonz\'{a}lez-F\'{e}rez}
\email{rogonzal@ugr.es}
\affiliation{Instituto 'Carlos I' de F\'{\i}sica Te\'orica y Computacional
and Departamento de F\'{\i}sica At\'omica, Molecular y Nuclear,
Universidad de Granada, E-18071 Granada, Spain}
\author{I. Lesanovsky}
\email{igor.lesanovsky@uibk.ac.at}
\affiliation{Institute of Theoretical Physics, University of Innsbruck and Institute for Quantum Optics and Quantum Information of the Austrian Academy of Sciences, Innsbruck, Austria}

\begin{abstract}
We study the excitation dynamics of Rydberg atoms in a one-dimensional lattice
with periodic boundary conditions where the atomic Rydberg states are
resonantly excited from the electronic ground state. Our description of the
corresponding dynamics is numerically exact within the perfect blockade regime,
i.e. no two atoms in a given range can be excited. The time-evolution of the mean Rydberg density, density-density correlations as well as entanglement properties are analyzed in detail. We demonstrate that the short time dynamics is universal and dominated by
quantum phenomena, while for larger time the characteristics of the lattice
become important and the classical features determine the dynamics.
The results of the perfect blockade approach are compared to the predictions
of an effective Hamiltonian which includes the interaction of two
neighboring Rydberg atoms up to second order perturbation theory.
\end{abstract} \maketitle

\section{Introduction}
During recent years, a new class of experiments in ultracold gases emerged, dedicated to the study of atoms excited to Rydberg states \cite{Gallagher} which interact strongly via dipole-dipole or van-der-Waals forces. The most intriguing manifestation of this interaction is the Rydberg blockade mechanism \cite{Jaksch00,Lukin01} which prevents the excitation of a Rydberg atom in the vicinity of an already excited one. On the theory side, this blockade has been thoroughly studied in the context of quantum information processing \cite{Jaksch00,Lukin01} for it is a natural implementation of a state dependent interaction which is essential to devise two qubit gates. Moreover, it has been theoretically shown that the long-ranged character of the interaction can be employed to manipulate whole atomic ensembles by just a single control atom \cite{Mueller08}.
Very recently, the state-dependent dynamics between two Rydberg atoms, spatially separated by several micrometers, was observed experimentally \cite{Urban08,Gaetan08}.

In the context of gases, a first experimental indication of the strong Rydberg-Rydberg interaction was the non-linear behavior of the number of excited atoms as a function of increasing laser power and atomic density \cite{Singer04,Tong04}. Later, it was shown that, in the case of a dense gas, the Rydberg blockade gives rise to the formation of coherent collective excitations - so-called 'superatoms' \cite{Heidemann07,Heidemann08}.
Due to the strong interaction, the timescale of the excitation dynamics of a Rydberg gas is typically much shorter than those of the external dynamics. As a consequence, such gases can be considered 'frozen' \cite{Mourachko98,Anderson98} in a given configuration and the evolution of Rydberg excitations is usually described by a spin model \cite{Sun08,Weimer08}, where the spin up/down state represents a Rydberg/ground state atom. Unlike in a typical solid state system, there are no significant dissipative processes which make the system assume its ground state over the typical experimental timescale. Therefore, one can regard the dynamics as fully coherent \cite{Raitzsch08,Reetz-Lamour08} and the time-evolution of quantities like the mean number of Rydberg excitations is expected to depend crucially on the initial state. An experiment, however, consists of many successive measurements which are performed for different configurations (initial states) of the atomic gas, i.e. changing positions of the atoms. Hence, the results of many experimental realizations have to be averaged in order find expectation values of physical quantities.

In the present work we study the excitation properties of a Rydberg gas in a particularly structured and symmetric scenario. In our setup, a large number $N_0\gg1$ of ground state atoms per site are homogeneously distributed over a ring lattice \cite{Amico05}. At most a single Rydberg atom per site can be excited via a resonant laser that is switched on instantaneously. Due to the underlying lattice the atomic configuration remains unchanged for any experimental realization and no averaging in the above sense is required. The atomic motion is considered frozen on relevant time scale. In the framework of the perfect blockade regime, we study the temporal evolution of the Rydberg excitation number, the formation of correlations in the Rydberg density and the entanglement properties in lattices with up to 25 sites.
We demonstrate that the dynamics of this system is divided into short and long time domains which are, respectively, independent and dependent on the lattice size.
Our calculations are numerically exact in the limit of the perfect Rydberg blockade and therefore might serve as reference for numerical methods that are developed for treating effectively one-dimensional many-particle systems. Effects that go beyond this regime are treated by second order perturbation theory, i.e. by a Hamiltonian which is obtained by adiabatically eliminating highly excited energy levels.

The paper is structured as follows: In Sec. \ref{sec:Hamiltonian} we derive an effective Hamiltonian which describes the Rydberg excitation dynamics of atoms confined to a ring lattice. Section \ref{sec:Symmetries} is dedicated to the discussion of the symmetry properties of the system and the consequent arising simplification of the numerical calculation. The regime of the perfect blockade is thoroughly studied in Sec. \ref{sec:Perfect}. In Sec. \ref{sec:Non-perfect} we discuss the adiabatic elimination procedure which we use to account for effects that go beyond the perfect blockade and perform a comparison of the results obtained by both approaches. The conclusion and outlook are provided in Sec. \ref{sec:Conclusions}.

\section{The Hamiltonian}\label{sec:Hamiltonian}
We consider a gas of bosonic atoms that is confined to a large spacing optical ring lattice with $N$ sites with periodicity $a \sim\mu m$. Such lattices have been proposed to be created by means of two interfering laser beams with $N$ being typically of the order of 20 \cite{Amico05}. Moreover, there is also a way to create a ring lattice approximatively in a standard rectangular large spacing optical lattice \cite{Nelson07} which is generated by crossed laser beams. To this end one can remove atoms from unwanted sites by employing the electron beam technique presented in Ref. \cite{Gericke08}, thereby 'cutting out' a ring.

For the sake of simplicity, we assume an uniform and large atomic density, i.e. the same number of ground state atoms $N_0\gg1$ per site. The atoms populate the ground state of each lattice site, which for the $k$-th one is described by the Wannier function $\Psi_k(\mathbf{x})$, where $\mathbf{x}$ represents the spatial coordinates. We assume no hopping and hence no particle exchange between the sites. The atoms are modeled as a two-level system: the ground state $\left|g\right>$ and the Rydberg state $\left|r\right>$. Experimentally these two levels are usually coupled by a two-photon transition. Here, we assume, without any loss of generality, that they are coupled resonantly by a laser of Rabi frequency $\Omega_0$. The corresponding Hamiltonian reads
\begin{equation}
H_0=\hbar\Omega_0\sum_{k=1}^N (b^{\dagger}_kr_k+b_kr^{\dagger}_k),\label{eqn:H0}
\end{equation}
where $b^{\dagger}_k$ and $r^{\dagger}_k$ ($b_k$ and $r_k$) represent the creation (annihilation) of a ground and a Rydberg state atom at the $k$-th site, respectively.
For the sake of simplicity we assume that the Rydberg atoms experience the same trapping potential than the ground state atoms. This requirement is, however, not crucial as the typical timescale of the electronic excitation dynamics which we are going to study is much smaller than the dephasing time due to different trapping potentials experienced by $\left|g\right>$ and $\left|r\right>$. Since we are working in the limit of a large number of ground state atoms, $N_0\gg 1$, contained in each site, they can be treated as a classical field and we can replace $b^{\dagger}_k$ and $b_k$ by $\sqrt{N_0}$. This yields the Hamiltonian
\begin{equation}
H_0=\hbar\Omega_0\sqrt{N_0}\sum_{k=1}^N (r_k+r^{\dagger}_k),\label{eqn:H0v2}
\end{equation}
where we assume a relative phase equal to $0$ between the condensates confined to different sites. Moreover, we neglect radiative decay. This is justified because the typical timescale of the excitation dynamics is given by the inverse of the collective Rabi frequency $\Omega\equiv\Omega_0\sqrt{N_0}$ which can easily exceed tens of MHz. We will later see that due to the Rydberg blockade at most $N/2$ Rydberg atoms will be excited on the ring. As a consequence, the time for the first emission of a photon scales in the worst case like $t_\mathrm{emit}=(2/N) T_0$, with the single Rydberg atom decay rate $T_0^{-1}\sim 100\, \mathrm{kHz}$. Thus, for the ring sizes under consideration the condition $\Omega \gg t^{-1}_\mathrm{emit}$ can be met.

Let us now discuss the interaction between the Rydberg atoms. We focus here on the van-der-Waals interaction which is given as $V(x)=C_6/x^{6}$ with $x$ being the interparticle distance. It has been shown \cite{Heidemann07,Heidemann08} that even this short-ranged, i.e. quickly decaying, interaction can significantly affect the excitation dynamics of atoms which are several $\mu$m apart. This is rooted in the large value of the coefficient $C_6$, which grows proportional to $n^{11}$ with $n$ being the principal quantum number of the excited level \cite{Singer05}. The interaction Hamiltonian is given by
\begin{eqnarray}
  H_\mathrm{int}=\sum_k V_{kk}n_k(n_k-1)+\frac{1}{2}\sum_{i\neq j} V_{i j} n_i n_j
\end{eqnarray}
where $n_k=r^\dagger_k r_k$ is the Rydberg particle number operator and $V_{ij}=\int \mathrm{d}\mathbf{x} \mathrm{d}\mathbf{x}^\prime \left|\Psi_i(\mathbf{x})\right|^2 \left|\Psi_j(\mathbf{x}^\prime)\right|^2 V(|\mathbf{x}-\mathbf{x}^\prime|)$.
We assume that the spatial extension of the Wannier functions $\Psi_j(\mathbf{x})$ is much smaller than the lattice spacing $a$. In this case we can write $V_{ij}\approx C_6/(|i-j|a)^6$, where we make the assumption that the range of interactions is much smaller than the radius of the ring lattice. Moreover, we assume that the on-site interaction is much larger than any other in the system, i.e. $\left| V_{kk}\right|\gg\left| V_{kk\pm1}\right|$. This implies that a double occupancy of a single site is ruled out and hence $n_k$ has the two eigenvalues $0$ and $1$. Within these approximations and after introducing the energy scale $\epsilon=\hbar\Omega$ the Hamiltonian reads
\begin{eqnarray}
  H=H_0+H_{\mathrm{int}}=\epsilon\sum_{k=1}^N\left[ (r_k+r^{\dagger}_k)+\sum_{l=1}^{m-1}\Delta_l n_k n_{k+l}\right]\label{eq:working_hamiltonian}
\end{eqnarray}
with $\Delta_l=C_6/[(l\,a)^6\,\epsilon]$. The maximal range of the interactions is given by $m\,a$, with $m$ a positive integer. The value of $m$ depends on the features of the considered system, such as the lattice spacing and the strength of the interaction. If the interaction rapidly decays as the distance is enhanced only a few neighboring sites interact. Since for the van-der-Waals potential $\left|V_{k k\pm2}\right|$ is 64 times smaller than $\left|V_{k k\pm1}\right|$, our study is focused on the $m=2$ case.

Without the possibility of a double occupancy, the operator $r_k^\dagger$ can be interpreted to create a superatom on the $k$-th site, i.e. a symmetric superposition of all possible single atom excitations on that site. This is done with a rate $\Omega$. In this spirit, the Hamiltonian (\ref{eq:working_hamiltonian}) describes the local dynamical creation and annihilation of such superatoms and their interaction. This Hamiltonian can be equivalently formulated as a spin model as it is done in Refs. \cite{Sun08,Weimer08}.

\section{Symmetries and state representation}\label{sec:Symmetries}
Our goal is to study the dynamics of this system with the vacuum state $\left|0\right>$ (such that $r_k\left|0\right>=0$) serving as initial state. To that purpose, we have to solve the time-dependent Schr\"odinger equation of the Hamiltonian (\ref{eq:working_hamiltonian}) and we perform this task by making use of a basis composed by all possible configurations in the ring. For an increasing site number $N$, the exact solution of this problem becomes quickly intractable as the dimension of total the Hilbert grows as $2^N$. By exploiting the symmetries of the Hamiltonian (\ref{eq:working_hamiltonian}) we can show, however, that the dimension of the subspace in which the evolution takes place is significantly reduced.

There are two symmetry operations on a ring lattice which are of interest in our system: cyclic shifts by $l$ sites and the reversal of the order of the lattice sites. The former is represented by the operators $X_l$ with $l=1,2,\dots,N$ where $X_l=X_1^l$, while the latter operation is the parity and is denoted by $R$. The action of these operations on the creation and annihilation operators is defined through
\begin{eqnarray}
X_l^{-1} r_k X_l=r_{k+l}&&
X_l^{-1}r_k^{\dagger}X_l=r_{k+l}^{\dagger}\\
R^{-1}r_kR=r_{N-k+1}&&
R^{-1}r_k^{\dagger}R=r_{N-k+1}^{\dagger},
\end{eqnarray}
from which follows that $R$ and all $X_l$ are unitary, i.e. $R^{-1}=R^\dagger$ and $X_l^{-1}=X_l^\dagger$. The Hamiltonian (\ref{eq:working_hamiltonian}) is invariant under these operations, i.e. $\left[H,X_l\right]=\left[H,R\right]=0$.

Let us now focus on the vacuum state $\left|0\right>$, where no Rydberg atom is excited. This is an eigenstate of \textit{all} cyclic shifts and the reversal operator with eigenvalue 1:
\begin{eqnarray}
  X_l\left|0\right>=\left|0\right>&&   R\left|0\right>=\left|0\right>.
\end{eqnarray}
Only a small subset of the $2^N$ states, spanning the whole Hilbert space, actually has these properties. Each of these maximally symmetric states can be understood as a superposition of all states that are equivalent under rotation and reversal of the sites.
Since the Hamiltonian conserves the symmetries of the initial state, the evolution of the system will take place in the subspace of the Hilbert space spanned by the maximally symmetric states. By using these states, the dimension of the problem is dramatically reduced. For example, for $N=10$ the dimension of the basis decreases from $2^{10}=1024$ down to $78$. For our computations we need an algorithm to quickly generate the maximally symmetric states among which the evolution takes place. Such an algorithm is presented in Ref. \cite{Sawada01}. There, these states are called bracelets,  and are recursively generated in an optimal way. The amount of CPU time grows only proportional to the number of bracelets produced.

\section{Perfect blockade}\label{sec:Perfect}
We will consider from now on the case $m=2$ if not explicitly said otherwise, i.e. only neighboring sites interact. The Hamiltonian (\ref{eq:working_hamiltonian}) is then governed by two energy scales: the one associated to the laser excitation $H_0$, i.e. the collective Rabi frequency $\epsilon=\hbar\Omega$ and the one related to the Rydberg-Rydberg interaction $H_\mathrm{int}$, given by $\epsilon \Delta \equiv \epsilon \Delta_1$.

\begin{figure}
\includegraphics[width=5cm]{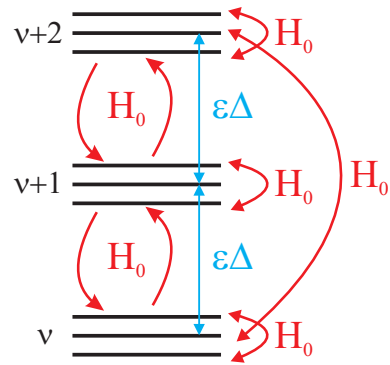}
\caption{Energy level structure of the Hamiltonian (\ref{eq:working_hamiltonian}) for $m=2$. The spectrum consists of highly degenerate subspaces which are labeled by $\nu$ energetically separated by $\epsilon\Delta$. The laser ($H_0$) causes an energy splitting of the degenerate levels. In addition, it couples states belonging to a given $\nu$-subspace and connects subspaces with $\vert\nu-\nu^\prime\vert=1,2$.}\label{fig:spectrum}
\end{figure}
The spectrum of the interaction Hamiltonian $H_\mathrm{int}$ decomposes into $N+1$ degenerate subspaces of energy $E_\nu=\nu\,\epsilon\Delta$ with $\nu=0,...,N$ counting the number of pairs of neighboring excitations. The laser Hamiltonian $H_0$ is switched on instantaneously and drives the dynamics within a given $\nu$-subspace and couples subspaces with $|\nu-\nu^\prime|\le2$ (see Fig. \ref{fig:spectrum}). The timescale associated with the evolution inside a $\nu$-subspace is $\tau_0=\hbar/\epsilon$, whereas the typical time of inter-subspace transitions is given by $\tau_\mathrm{int}=\hbar\Delta/\epsilon=\tau_0\Delta$. We consider here the regime in which $\Delta\gg 1$, i.e., the interaction energy of two neighboring Rydberg atoms is much larger than the collective Rabi frequency and hence $\tau_0\ll\tau_\mathrm{int}$. For the van-der-Waals case, this condition is $\hbar\Omega\ll \vert C_6\vert/a^6$. %$\mathbf{a=400\,nm,\, n=40,\,\Omega_0=10\,MHz,\,N_0=100.}$

The physical initial state is the vacuum, $\vert0\rangle$ which belongs to the subspace with $\nu=0$. In our approximation we neglect the coupling between this subspace and those including higher excitations. This is the perfect blockade approach, which is valid for $t\ll\tau_\mathrm{int}$, i.e. the time it takes to perform a transition between adjacent $\nu$-subspaces. The restriction to the $\nu=0$ subspace leads to a further reduction of the dimension of Hilbert space in which the temporal evolution takes place. For example, for $N=10$ the number of states to be considered in the basis set expansion decreases from $78$ to $14$. This is to be compared to the $1024$ states which span the entire Hilbert space of the system.

\begin{figure}
\includegraphics[width=8.5cm]{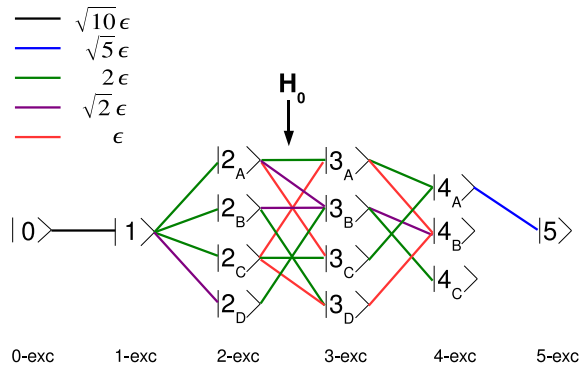}
\caption{Graph with the basis of states for $N=10$ in which the time-evolution takes place. In each column a subspace of a given number of Rydberg excitations is shown, see text for further information. The laser (Hamiltonian $H_0$) couples only states belonging to adjacent subspaces. The coupling strength (transition probability) between the individual states is encoded in the colors.}\label{fig:graph}
\end{figure}

The laser Hamiltonian $H_0$ couples states in the Hilbert space whose number of excitations differ by one. The corresponding maximally symmetric states and the coupling between them by means of $H_0$ can be graphically illustrated as shown in Fig. \ref{fig:graph}.
The way these states are coupled is qualitatively similar for different lattice sizes so, for simplicity reasons, we discuss here the lattice with $N=10$.
In Fig. \ref{fig:graph} the states are denoted by the number of Rydberg excitation, and a subscript is added when more than one configuration with the same excitation number is possible. Note that, for an even (odd) number of sites the maximal number of Rydberg atoms in one of these states is $\frac{N}{2}$ ($\frac{N-1}{2}$), e.g. $|5\rangle$ for $10$ sites.

Starting from the vacuum, there are several excitations paths with different probabilities that connect the states. The larger the amount of Rydberg atoms, the more constraints are found to allocate the next excitation. As a consequence, we encounter several excitation paths that do not reach the state with the maximal number of Rydberg excitations, but end in others, such as the $|4_B\rangle$ and $|4_C\rangle$ states for $N=10$. The features of these frustrated states strongly depend on the lattice size, and their amount increases as $N$ is increased. In particular, their existence provokes quantitative differences in the dynamics of two lattices with different $N$-value for large times.
This is reflected in the time-evolution of all quantities we are going to study throughout this work. The dynamics can be always divided into two different domains. For short times, $t\lesssim4\tau_0$ the behavior is universal, i.e., independent of the size of the lattice, whereas for longer times a dependence on $N$ is observed.

\subsection{Two-sites density matrix}
The reduced density matrix of two neighboring sites is needed to investigate the temporal evolution of local properties such as the mean density of Rydberg atoms or the entanglement between two adjacent sites. Since the wavefunction $\left|\Psi(t)\right>$ is spanned in the subspace of fully symmetric states, all sites are indistinguishable and we can take $1$ and $2$ as representative adjacent lattice sites. The two-sites reduced density matrix is obtained from the density matrix of the full system $\rho(t)=\vert\Psi(t)\rangle\langle\Psi(t)\vert$ by performing the partial trace over all the remaining sites,
\begin{equation}
\rho^{(12)}(t)=\mathrm{Tr}_{3,4\dots N}\left(\rho(t)\right).
\end{equation}
The basis for the two-sites states is $\{\vert gg \rangle, \vert gr \rangle, \vert rg \rangle, \vert rr \rangle\}$. The restriction to the fully symmetric subspace imposes  $\langle gr\vert\rho^{(12)}\vert gr\rangle=\langle rg\vert\rho^{(12)}\vert rg\rangle\equiv\beta$ and $\langle gg\vert\rho^{(12)}\vert gr\rangle=\langle gg\vert\rho^{(12)}\vert rg\rangle\equiv\gamma$, while the perfect blockade prevents the excitation of atoms in two neighboring sites and hence the entries $\langle A\vert\rho^{(12)}\vert rr\rangle$ and $\langle rr\vert\rho^{(12)}\vert A\rangle$ are zero for $\vert A \rangle=\{\vert gg \rangle, \vert gr \rangle, \vert rg \rangle\}$.
As a consequence, the reduced density matrix has the particularly simple form:
\begin{equation}\label{eqn:dens2}
\rho^{(12)}(t) =
\left( \begin{array}{cccc}
\alpha & \gamma & \gamma & 0 \\
\gamma^* & \beta & \delta & 0 \\
\gamma^* & \delta^* & \beta & 0 \\
0 & 0 & 0 & 0
\end{array} \right),
\end{equation}
where $\alpha$, $\beta$, $\gamma$ and $\delta$ are four time-dependent (complex) parameters. In particular, $\alpha$ and $\beta$ are real, and due to the normalization of the wavefunction, $\mathrm{Tr}_{1,2}\,\rho^{(12)}(t)=\alpha+2\beta=1$. Hence, only three of these parameters are independent. Performing the trace over the states of site number 2 yields the single particle density matrix
\begin{equation}
\rho^{(1)}(t)=\mathrm{Tr}_2\left(\rho^{(12)}(t)\right) =
\left( \begin{array}{cc}
1-\beta & \gamma \\
\gamma^* & \beta
\end{array} \right).\label{eq:single_site_DM}
\end{equation}

\subsection{Rydberg density}
\begin{figure}
\includegraphics[width=8.5cm]{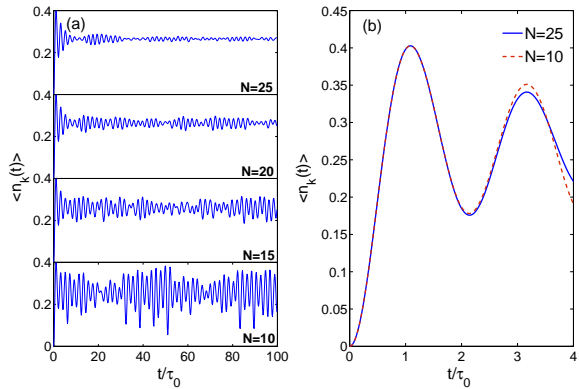}
\caption{Expectation value of the Rydberg density (\ref{eqn:Ryd_dens}) versus time for (a) four different ring sizes and $t\leq100\,\tau_0$, and (b) detail of the short time evolution for $N=10\,\mathrm{and}\,25$. The computations have been performed assuming perfect blockade of the adjacent neighbor.}\label{fig:NRydvsT}
\end{figure}

The first local property under consideration is the time evolution of the expectation value of the Rydberg density $n_k$. By using Eq. (\ref{eq:single_site_DM}) one obtains
\begin{equation}\label{eqn:Ryd_dens}
\langle n_k(t)\rangle=\mathrm{Tr}\left(\rho^{(1)}(t)n_k(t)\right)=\beta,
\end{equation}
and the total number of Rydberg atoms evaluates to
$\langle N_\mathrm{Ryd}(t)\rangle=N\,\langle n_k(t)\rangle=N\beta$. In Fig. \ref{fig:NRydvsT}a we show $\langle n_k(t)\rangle$ as a function of time for different lattice sizes. In the right panel (Fig. \ref{fig:NRydvsT}b) a magnified view of the short time dynamics is provided for $N=10$ and $25$.

We observe at first a steep increase, which is proportional to $t^2$, that culminates in a pronounced peak located at $t=1.09\,\tau_0$. This peak is independent of the lattice size, as we are still witnessing the short time behavior. For much larger times, $\langle n_k(t)\rangle$ becomes dependent of $N$ and oscillates with a frequency $f \approx 0.48\, \Omega$ about a mean value of $\langle\overline{n_k(t)}\rangle\approx 0.26$. This mean value and also $f$ turn out to be independent of the ring size, however, the exact shape of $\langle n_k(t)\rangle$ strongly depends on $N$. A similar result and a possible explanation of the nature of this value is given in Ref. \cite{Sun08}. The time averages are performed by means of numerical integration over time in a large enough interval $\left[5\tau_0,200\tau_0\right]$. The lower limit of this interval is chosen large enough to avoid the initial effects of turning on the laser. The higher one is kept shorter than the corresponding revival time. The amplitude of the oscillations decreases considerably with increasing lattice size. This amplitude can be measured by means of the standard deviation. For $N=10$, a quasi-steady state with large fluctuations characterized by a standard deviation $\sigma(\langle n_k(t)\rangle)=0.062$ about $\langle \overline{n_k(t)}\rangle$ is established, while in the case of $N=25$ these fluctuations become smaller $\sigma(\langle n_k(t)\rangle)=0.0065$.

We have also performed calculations of time-averaged Rydberg density $\langle\overline{n_k(t)}\rangle$ for $m=3$ and $m=4$ and $N>20$. The results for the two cases are $\langle\overline{n_k(t)}\rangle_{\mathrm{m=3}}\approx 0.17$ and $\langle\overline{n_k(t)}\rangle_{\mathrm{m=4}}\approx 0.12 $, which is an indication for the scaling $\langle\overline{n_k(t)}\rangle_{\mathrm{m}}\approx (2\,m)^{-1}$.

\subsection{Density-density correlation function}
The equal-time density-density correlation function between two sites denoted by $i$ and $j$ separated by a distance $\left|i-j\right|\,a\equiv k\,a$ is given by
\begin{equation}
g_2(k,t)=\frac{\langle n_1n_{1+k}(t)\rangle}{\langle n_1(t)\rangle\langle n_{1+k}(t)\rangle}.
\end{equation}
It measures the conditional probability of finding an excited atom at a distance $k\,a$ from an already excited one normalized to the probability of uncorrelated excitation.

\begin{figure}
\begin{center}
\includegraphics[width=8.5cm]{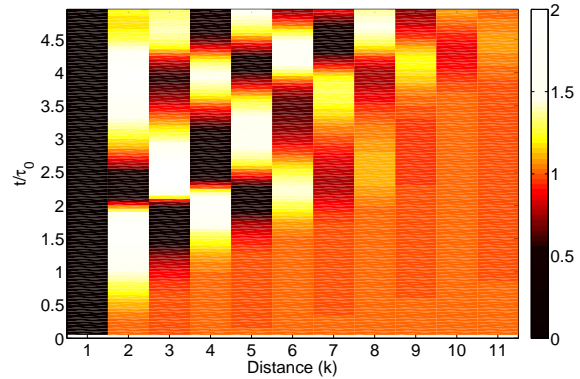}
\caption{Short time behavior of the density-density correlations $g_2(k,t)$ for $N=25$.
Correlations emerge successively during the time evolution. Due to the perfect blockade condition strong oscillations of $g_2(k,t)$ with a period $k=2$ are observed.}\label{fig:G2lowtime}
\end{center}
\end{figure}

Figure \ref{fig:G2lowtime} illustrates the initial evolution of $g_2(k,t)$ in the time interval $\left[0,5\,\tau_0\right]$ for a $N=25$ lattice.
Due to the perfect blockade condition, $g_2(1,t)=0$ for any time. The temporal and spatial structure can be understood by observing the properties of the laser Hamiltonian $H_0$ which drives the excitation dynamics.
At the beginning only a single particle at site 1 is excited, so the probability of excitation of a second atom is uniform, it can occur at arbitrary position (except at a distance of $k=1$). As a consequence, there are no correlations for very short times, and they emerge successively as time increases.
The augment occurs at $k=2$, and the density-density correlation function for that distance reaches a maximum at $t\approx 1.5\,\tau_0$. The high probability of finding two excitations at the distance $k=2$, i.e. a large value of $g_2(2,t)$, automatically gives rise to a decrease of $g_2(3,t)$ due to Rydberg blockade. For larger times, a regular pattern of enhanced and suppressed density-density correlations characterizes the dynamics. The regular pattern of the density-density correlation functions at short times is lost as time increases. Here, $g_2(k,t)$ exhibits pronounced fluctuations and rapid oscillations around a mean value for $t\gtrsim4\,\tau_0$.
\begin{figure}
\begin{center}
\includegraphics[width=8.5cm]{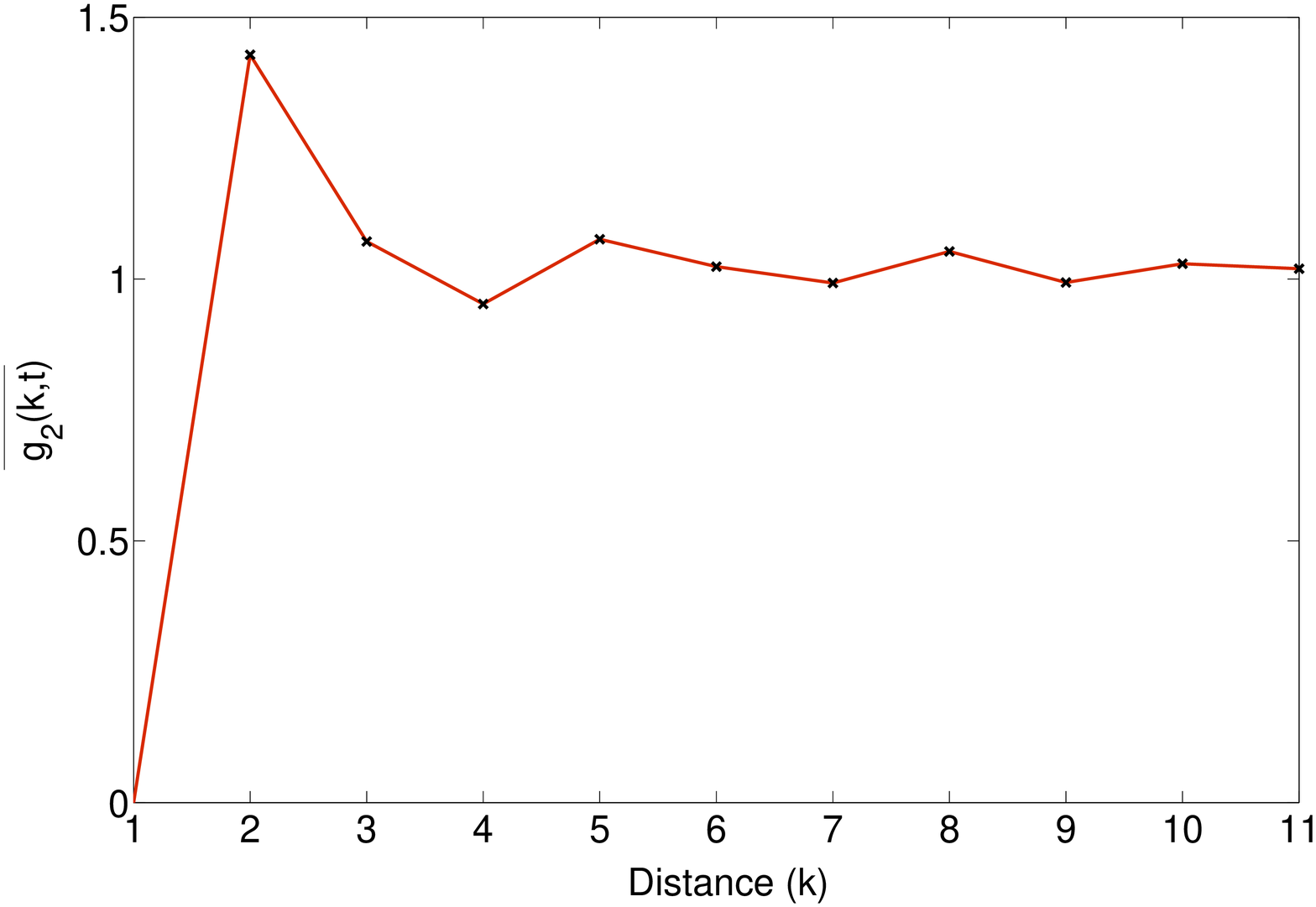}
\caption{Time-averaged density-density correlations $\overline{g_2(k,t)}$ for $N=25$. For distance $k=1$, the value is zero due to the perfect blockade. $\overline{g_2(k,t)}$ assumes a maximum for $k=2$ (next-nearest neighbor). For larger distances, only weak correlations are visible.}\label{fig:G2}
\end{center}
\end{figure}

In Fig. \ref{fig:G2} we show the time-averaged density-density correlation function, $\overline{g_2(k,t)}$, in the stationary long time regime as a function of $k$, for a lattice of $25$ sites.
This function shows a maximum for $k=2$, while for larger intersite distances it approaches the constant value 1, i.e., no correlations. As a consequence, we conclude that the density-density correlations are only short ranged after the initial period in which also long ranged correlations are of importance.

\subsection{Entanglement}
We study the quantum and classical correlations and the entanglement of two
neighboring sites in this system by means of the \emph{two-party correlation
  measure} \cite{Zhou06} and the \emph{entanglement of formation}
\cite{Wooters98}. These quantities can be directly related to the entries of
the reduced density matrix discussed previously.

\begin{figure}
\begin{center}
\includegraphics[width=8.5cm]{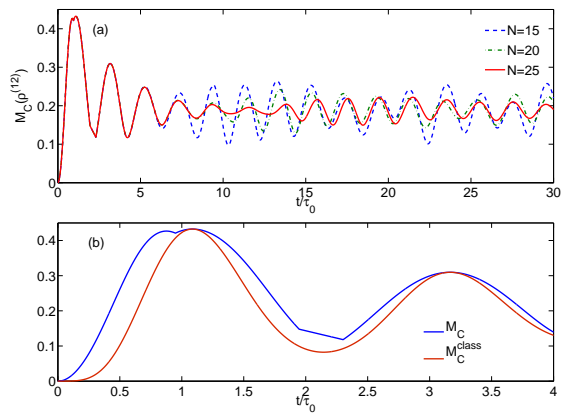}
\caption{(a) Time-evolution of the two-party correlation measure $M_C\left(\rho^{(12)}\right)$ for long times and various lattice sizes. (b) Short time behavior of $M_C\left(\rho^{(12)}\right)$ and its classical counterpart for $N=25$.}\label{fig:Totalcorr}
\end{center}
\end{figure}

\paragraph*{Two-party correlation.} The two-party correlation measure \cite{Zhou06} is based on the trace distance \cite{Nielsen} and it is defined as
\begin{equation}\label{eqn:total_corr}
M_C\left(\rho^{(12)}\right)=\frac{2}{3}\mathrm{Tr}\vert\rho^{(12)}-\rho^{(1)}\otimes\rho^{(2)}\vert,
\end{equation}
where $\vert A\vert\equiv\sqrt{A^{\dagger}A}$ is the positive square root of $A^{\dagger}A$. Its physical meaning is the distance between the state $\rho^{(12)}$ and its reduced product state $\rho^{(1)}\otimes\rho^{(2)}$. It takes into account both the classical correlation between two sites and the quantum coherence. It generalizes the classical distance in the sense that if the two operators commute then it is equal to the classical trace or Kolmogorov distance between the eigenvalues of $\rho^{(12)}$ and $\rho^{(1)}\otimes\rho^{(2)}$. We show the time evolution of this correlation measure in Fig. \ref{fig:Totalcorr}a for different sizes of the ring.
Initially, for the vacuum state, there are no correlations. Analogously to the previously analyzed quantities, $M_C$ exhibits an $N$-independent short time behavior which here is characterized by large amplitude oscillations. It is followed by an $N$-dependent regime, where $M_C$ presents smooth oscillations around the mean value $\overline{M_C}=0.19$. As expected, the amplitude of these oscillations decreases with increasing lattice size.

We are now interested in finding a classical counterpart to this correlation measure. To this end, we make use of the density matrix properties. The diagonal elements of a density matrix represent the probability of finding the corresponding configuration of the sites. For example, in $\rho^{(12)}$, (see equation (\ref{eqn:dens2})), $\alpha$ and $\beta$ represent the probability of the two sites to be in the states $\left|gg\right>$ and $\left|gr\right>$ or $\left|rg\right>$, respectively; note that the sites are
indistinguishable. In the same way, the diagonal components of the reduced product density matrix $\rho^{(1)}\otimes\rho^{(2)}$ provides the probability of the two sites being in the corresponding product state,  e.g. $(1-\beta)^2$ for the state $\left|g\right>\otimes\left|g\right>$.

We take these diagonal elements $d^{(12)}_i$ and $d^{(1\otimes2)}_i$ of the matrices $\rho^{(12)}$ and $\rho^{(1)}\otimes\rho^{(2)}$, respectively, as classical probability distributions. The Kolmogorov distance between these distributions, is defined here as:
\begin{equation}
M_C^\mathrm{class}\left(\rho^{(12)}\right)\equiv\frac{2}{3}\sum_{i=1}^{4}\vert d^{(12)}_i-d^{(1\otimes2)}_i\vert,
\end{equation}
and it provides a classical measure of the two-party correlation. In terms of the parameters of the density matrix this quantity is reduced to
\begin{equation}
M_C^\mathrm{class}\left(\rho^{(12)}\right)=\frac{8}{3}\beta^2.
\end{equation}
The classical and the total two-party correlation functions are presented in Fig. \ref{fig:Totalcorr}b for $t\leq4\,\tau_0$ and 25 sites.
One of the main features due to the quantum behavior of the system is the appearance of the two consecutive peaks of $M_C$ at $t=0.88\,\tau_0$ and $t=1.09\,\tau_0$. Note that the classical counterpart $M_C^{\mathrm{class}}$ reproduces only the second maximum. Hence, the existence of the first one can be only justified by quantum arguments.
Due to the absolute values in expression (\ref{eqn:total_corr}), two discontinuities appear in the derivative of $M_C$ around $t\approx1.9\,\tau_0$ and $t\approx2.3\,\tau_0$. They are, however, not observed in $M_C^{\mathrm{class}}$ since it only depends on the single, smoothly varying, parameter of the density matrix, $\beta$.

\begin{figure}[h]
\begin{center}
\includegraphics[width=8cm]{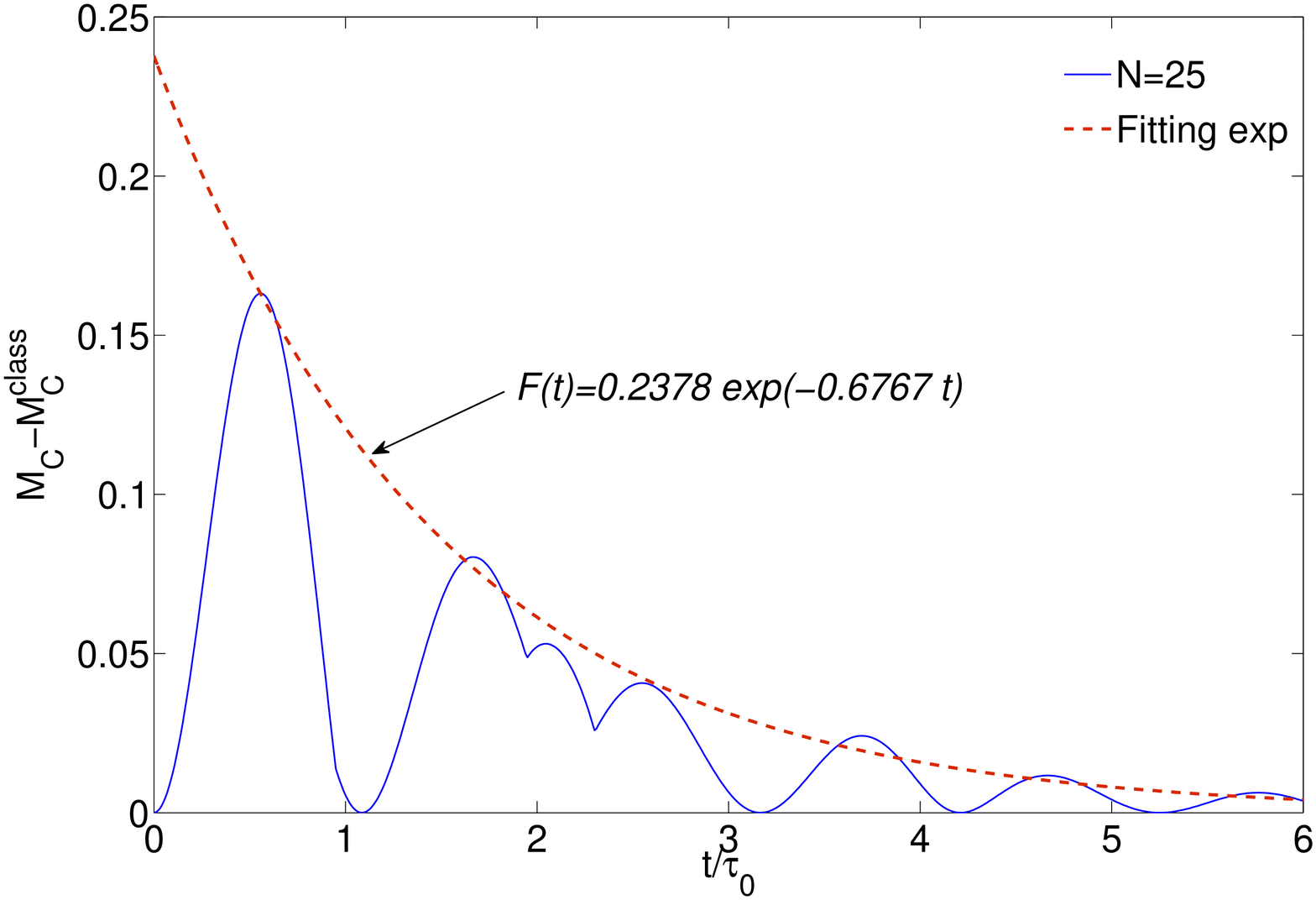}
\caption{Difference between the two-party correlation measure and its classical counterpart for $N=25$. The dashed line corresponds to an exponential fit to the envelope of this difference.}\label{fig:MC-MCclass}
\end{center}
\end{figure}

To get a deeper insight into the quantum effects on the correlations, the difference between the total two-party correlation and the classical measure is shown in Fig. \ref{fig:MC-MCclass} as a function of time. We have performed a fit to the local maxima of this numerical difference using an exponential decreasing function. The contribution of the quantum correlations loses importance as time is increased and, at the same time, the classical dynamics starts to dominate the correlations between two neighboring sites.

\paragraph*{Concurrence and entanglement of formation.} For a general state of two qubits represented by means of its two-particles density matrix $\rho$, the concurrence is given by \cite{Wooters98}
\begin{equation}\label{Concurrence}
C(\rho)=\mathrm{max}\left\{0,\lambda_1-\lambda_2-\lambda_3-\lambda_4\right\},
\end{equation}
where the $\lambda_i$ are the square roots of the eigenvalues, in decreasing order, of the matrix $\rho\tilde{\rho}$, where $\tilde{\rho}$ is the flipped matrix of the two-qubit general state $\rho$, i.e.,
\begin{equation}
\tilde{\rho}^{(12)}=\left(\sigma_y\otimes\sigma_y\right)\left(\rho^{(12)}\right)^*\left(\sigma_y\otimes\sigma_y\right).
\end{equation}

The entanglement of formation of a state of two qubits is defined as
\begin{equation}
E(\rho)=h\left(\frac{1+\sqrt{1-C(\rho)^2}}{2}\right),
\end{equation}
with $h(x)=-x\log_2x-(1-x)\log_2(1-x)$.
This quantity provides a measure of the resources needed to create a certain entangled state, and its range goes from 0 to 1.

Using the two-sites density matrix describing our system (\ref{eqn:dens2}), and its corresponding flipped matrix, we obtain the following $\lambda_i$:
\begin{equation}
\lambda_1=(\beta+|\delta|);\qquad\lambda_2=\vert\beta-|\delta|\vert;\qquad\lambda_3=\lambda_4=0.
\end{equation}
These values give rise to two different regimes for the concurrence:
\begin{equation}
C\left(\rho^{(12)}\right)=\left\{\begin{array}{cc} 2\vert\delta\vert & \beta>\vert\delta\vert \\ 2\beta & \beta<\vert\delta\vert \end{array}\right..
\end{equation}
The first condition $\beta>\vert\delta\vert$ always holds for any size $N$ of the lattice, so the concurrence yields
\begin{equation}
C\left(\rho^{(12)}\right)=2\vert\delta\vert.
\end{equation}

\begin{figure}
\begin{center}
\includegraphics[width=8.5cm]{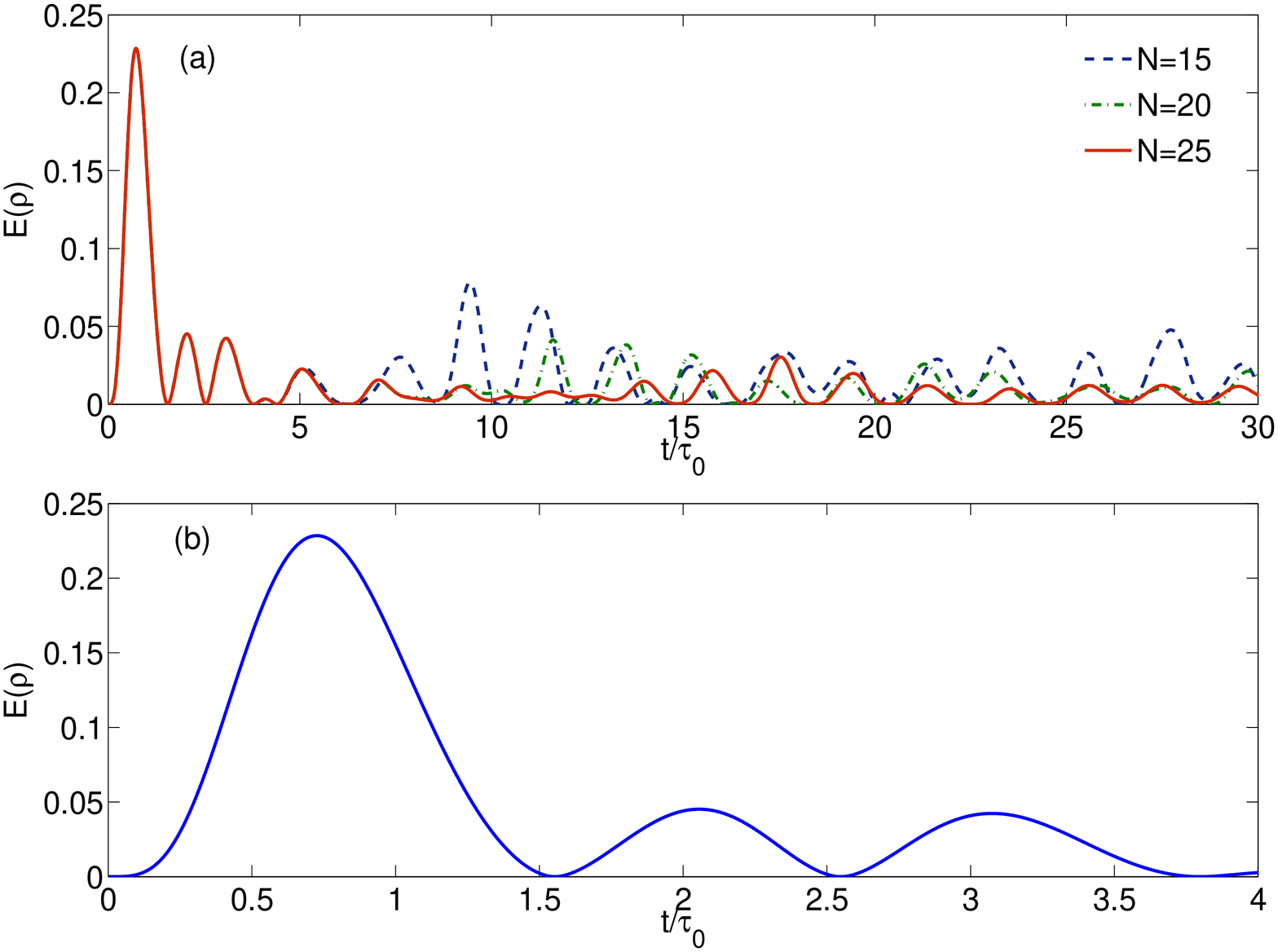}
\caption{Time-evolution of the entanglement of formation for (a) long time and several sizes of the system  and (b) short time for $N=25$.}\label{fig:eof}
\end{center}
\end{figure}

The time evolution of the entanglement for the lattices with sites $N=15,20$ and $25$ is presented in Fig. \ref{fig:eof}a, and an enhancement of the behavior at short times for $N=25$ is shown in Fig. \ref{fig:eof}b. Again, two different time domains can be distinguished. For short times, the entanglement of formation is independent of the ring size. Its maximal value, $E\left(\rho\right)=0.23$, is reached at $t=0.73\,\tau_0$; for a further increase of time, $E\left(\rho\right)$ drastically decreases, e.g. the second peak at $t=2.05\,\tau_0$ is reduced roughly by $80\%$. In the long time regime, the entanglement becomes weaker with $E\left(\rho\right)$ eventually approaching zero with characteristic fluctuations for each $N$. The amplitudes of these fluctuations become smaller as $N$ is increased.

\section{Non-perfect Blockade}\label{sec:Non-perfect}
The above discussed phenomena have been investigated in the perfect blockade regime, which assumes that the energy scale associated to the van-der-Waals interaction is infinitely large compared to the one related to the laser interaction, i.e. $\Delta\gg1$. Since the initial state has no Rydberg excitations, the dynamics of the system is restricted to a small subspace including those eigenstates of $H_\mathrm{int}$ with eigenvalue zero. However, for finite values of $\Delta$, the states with $\nu=0$ are coupled to those with $\nu>0$ and, even more, these higher excitations might influence the dynamics.

In this section, we go beyond the perfect blockade approach and explore these couplings including their effect of up to order $1/\Delta$. We thereby derive an effective Hamiltonian by dividing the eigenstates of $H_\mathrm{int}$ into two sets, characterized by their respective quantum number $\nu$. The first set of states is formed by the subspace $\nu=0$, whereas the second one contains the rest of energetically high-lying excitations. In this framework, the Hamiltonian can be written as
\begin{equation}
H\equiv\left(\begin{array}{cc}
PHP & PHQ \\
QHP & QHQ
\end{array}
\right),
\end{equation}
where $P$ and $Q$ are the projectors on the subspaces with energy $\nu=0$ and $\nu>0$, respectively. A general wavefunction can be decomposed as
\begin{equation}
\Psi\equiv\left(\begin{array}{c} P\Psi \\ Q\Psi \end{array}\right),
\end{equation}
and the time-dependent Schr\"{o}dinger equation reads:
\begin{equation}\label{eq:sch1_ad}
i\hbar\partial_t\left(\begin{array}{c} P\Psi \\ Q\Psi \end{array}\right)=\left(\begin{array}{cc}
PHP & PHQ \\
QHP & QHQ
\end{array}
\right)\left(\begin{array}{c} P\Psi \\ Q\Psi \end{array}\right).
\end{equation}
Due to the large energetic gap between the different subspaces and since the initial state is the vacuum, the transition probability to states with $E_\nu\ne E_0$ is very small. Thus, we can introduce an approximation assuming that the time variation of $Q\Psi$ is very small and can therefore be neglected, i.e. $\partial_t (Q\Psi)=0$. Hence, the equation of motion (\ref{eq:sch1_ad}) is reduced to
\begin{equation}\label{eqn:H_eff}
i\partial_t(P\Psi)=\left(PHP-PHQ(QHQ)^{-1}QHP\right)(P\Psi).
\end{equation}
Note that, in this expression, $PHP=H_0$ is the Hamiltonian within the perfect blockade regime. Whereas, the second term provides the first correction to this Hamiltonian and  represents the contribution of the couplings between the $\nu=0$ and $\nu>0$ subspaces.
In practice, $H_0$ only couples the $\nu=0$ subspace and those with $\nu=1\,\mathrm{and}\,2$, see Fig. \ref{fig:spectrum}. As a consequence, the Hamiltonian can be rewritten as
\begin{equation}\label{eq:H_ad}
H\equiv\left(\begin{array}{cccc}
H_0 & \Omega_{01} & \Omega_{02} & 0\\
\Omega_{10} & \epsilon\Delta+\Omega_1 & \Omega_{12} & \Omega_{1R}\\
\Omega_{20} & \Omega_{21} & 2\epsilon\Delta+\Omega_2 & \Omega_{2R}\\
0 & \Omega_{R1} & \Omega_{R2}& \Delta_R+\Omega_R
\end{array}
\right),
\end{equation}
where the subscript $R$ denotes the energetic levels with $\nu>2$, and the $\Omega_{ab}$ represent the part of the Hamiltonian that couples the states of the subspaces with  $\nu=a$ and $\nu^\prime=b$.
In expression (\ref{eq:H_ad}), $QHQ$ can be decomposed into the sum of a diagonal matrix, $\bar{\Delta}$, including the interaction between the subspaces, and a full matrix containing the couplings $\bar{\Omega}$,
\begin{eqnarray}\nonumber
QHQ&=&\left(\begin{array}{ccc}
\epsilon\Delta & 0 & 0\\
0 & 2\epsilon\Delta & 0\\
0 & 0 & \Delta_R
\end{array}
\right)+\left(\begin{array}{ccc}
\Omega_1 & \Omega_{12} & \Omega_{1R}\\
\Omega_{21} & \Omega_2 & \Omega_{2R}\\
\Omega_{R1} & \Omega_{R2}& \Omega_R
\end{array}
\right)\\
&\equiv&\bar{\Delta}+\bar{\Omega}.
\end{eqnarray}
The inverse of this matrix can be approximated by
\begin{equation}
(QHQ)^{-1}=\frac{1}{\bar{\Delta}+\bar{\Omega}}\approx\bar{\Delta}^{-1}-\bar{\Delta}^{-1}\bar{\Omega}\bar{\Delta}^{-1}+\dots,
\end{equation}
where we have used the Neumann series $\left(\mathbb{I}-T\right)^{-1}=\sum_{n=0}^{\infty}T^n$ for a square matrix $T$ whose norm satisfies that $\|T\|<1$.
Since $\Delta\gg\Omega_{ab}$ for any $a\,\mathrm{and}\,b$, this condition is accomplished for $T=\bar{\Delta}^{-1}\bar{\Omega}$.
Finally, we obtain the following expression for the effective Hamiltonian
\begin{equation}\label{eqn:H_effective}
H_\mathrm{eff}= H_0-\frac{\Omega_{01}\Omega_{10}}{\epsilon\Delta}-\frac{\Omega_{02}\Omega_{20}}{2\epsilon\Delta}+O(1/\Delta^2),
\end{equation}
where we only consider the first three terms and neglect higher order corrections.

Let us now discuss the regime of validity of the approximate Hamiltonian (\ref{eqn:H_effective}). To this end it is instructive to study a case in which the full Hamiltonian (\ref{eq:working_hamiltonian}) is numerically tractable. This however, can only be done for a small number of sites. In the absence of the laser the eigenstates of Hamiltonian (\ref{eq:working_hamiltonian}) are those of $H_\mathrm{int}$, i.e. the highly degenerate $\nu$-manifolds. As soon as the laser is turned on this degeneracy is lifted and all the $\nu$ manifolds split up. However, if $\Delta$ is sufficiently large the manifolds are still well separated.
\begin{figure}
\begin{center}
\includegraphics[width=8.5cm]{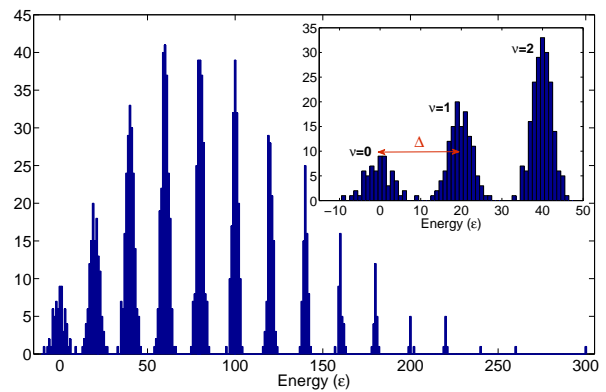}
\caption{Histogram of all the eigenvalues (density of states) of the full Hamiltonian for a system with $N=15$, $m=2$ and $\Delta=20$. The parameters are chosen such that the individual $\nu$-manifolds are still recognizable. The inset shows a magnified view of the manifolds with $\nu=0,1,2$, which are broadened by the interaction with the laser. }\label{fig:gap}
\end{center}
\end{figure}
This regime is presented in Fig. \ref{fig:gap} where we show a histogram of the eigenenergies (density of states) for a lattice with $N=15$ and $\Delta=20$. Since in this case the system can contain at most $15$ pairs of consecutive Rydberg atoms, we observe $16$ manifolds, i.e. $0\le\nu\le15$. The energetic separation between the central states of two neighboring subspaces is given by $\Delta$. A magnified view of the spectral structure for the low-lying excitations is shown in the inset of Fig. \ref{fig:gap}. Within the framework of the adiabatic elimination the contribution of the $\nu=1$ and $2$ manifolds is included up to order $\frac{1}{\Delta}$ in the effective Hamiltonian (\ref{eqn:H_effective}). The validity of this approximation is restricted to parameter regimes in which states belonging to different manifolds are energetically well-separated, e.g. two adjacent manifolds must not overlap. For $N=15$, $\Delta=20$ is the minimal value needed to ensure this separation. For larger lattices sizes, the value of $\Delta$ has to be increased since with growing $N$ the $\nu$-manifolds contain more and more states and thus become successively broader. For example, the width of the $\nu=0$ manifold scales proportional to $N$.

\begin{figure}
\begin{center}
\includegraphics[width=8.5cm]{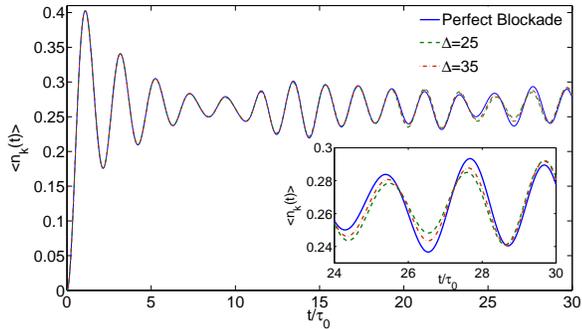}
\caption{The Rydberg density as a function of time for $N=20$ computed with the perfect blockade treatment, and with the adiabatic elimination scheme using $\Delta=25\,\mathrm{and}\,35$.}\label{fig:adiabatic}
\end{center}
\end{figure}

\begin{figure}
\begin{center}
\includegraphics[width=8.5cm]{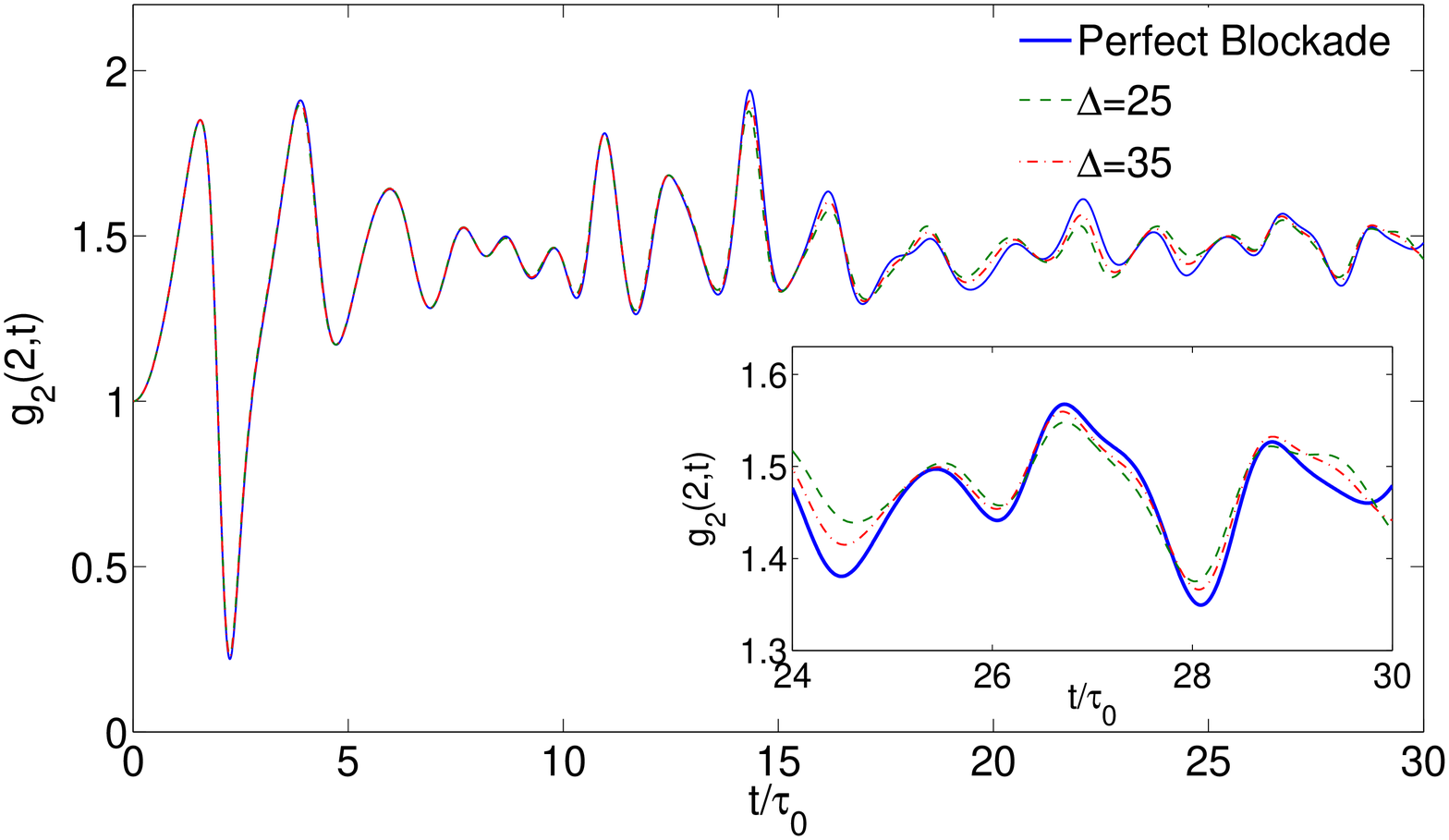}
\caption{Correlation function for $k=2$ as a function of time for $N=20$ computed with the perfect blockade treatment, and with the adiabatic elimination scheme using $\Delta=25\,\mathrm{and}\,35$.}\label{fig:g2_adiabatic}
\end{center}
\end{figure}

We have investigated the dynamics of a ring with $N=20$ sites in the framework of the adiabatic elimination using $\Delta=25$ and $35$. In Figs. \ref{fig:adiabatic} and \ref{fig:g2_adiabatic} we show the Rydberg density and a density-density correlation function (for $k=2$) and compare them with the results obtained within the perfect blockade approximation. The occurring deviations are small. Only minor differences are observed at large times, and we encounter relative errors below $6.5\%$ and $4\%$ for $\Delta=25$ and $35$, respectively.
The results show that the approximated inclusion of higher $\nu$-subspaces in the dynamics does only lead to small quantitative changes in the behavior of the investigated quantities. As anticipated, the deviations reduce significantly as $\Delta$ is increased. More qualitative differences are expected to occur if the $r^{-6}$-tail of the Rydberg-Rydberg interaction is properly accounted for.

\section{Conclusions and Outlook}\label{sec:Conclusions}

In this work we have performed a numerical analysis of the laser-driven Rydberg excitation dynamics of atoms confined to a ring lattice. By exploiting the symmetry properties of the system and employing the assumption of a perfect Rydberg blockade we were able to perform numerically exact calculations in lattices with up to N=25 sites. Our findings show that the temporal evolutions of the physical quantities, e.g. the Rydberg density and the density-density correlations, can be divided into two domains. For short times, one observes an N-independent universal behavior with large amplitude oscillations. For longer times, the dynamics is crucially determined by the lattice size and the analyzed quantities appear to assume a quasi steady state with only small temporal fluctuations. Moreover, we studied the evolution of the entanglement as well as the quantum and classical correlation of two neighboring sites. By separating the quantum and classical part of the two-party correlation we showed that quantum correlations between neighboring sites decay rapidly as time passes. In addition, the entanglement between neighboring sites turned out to be weak in the long time limit after a quick initial increase.

We eventually relaxed the perfect blockade condition by taking into account
higher excitation subspaces via adiabatic elimination. Propagating the initial
vacuum state with the corresponding effective Hamiltonian has only small
effect on the time evolution of the investigated quantities. More quantitative
changes are expected when including the long-ranged tails of the interatomic
interaction potential. This however requires more powerful numerical methods,
such as t-DMRG \cite{Vidal04}, which goes beyond the scope of this work.

In the present work we have been focusing on the dynamical properties of this system. A next step would be to investigate the corresponding static properties, such as eigenstates and eigenvalues. However, since physically the system is at $t=0$ in the vacuum state it remains an open question how the eigenstates can be actually accessed, for instance via an adiabatic passage incorporating a time-dependent change of the laser detuning and its Rabi frequency. In addition, a rather natural extension would be to analyze the dynamics of these many-particle systems by means of a two dimensional description. Certainly, it is also of interest to explore lattices with different geometries, e.g. square or triangle. The different underlying symmetries are expected to significantly affect the time evolution of these systems.

\begin{acknowledgments}
Financial support by the Spanish projects FIS2008-02380 (MEC) and FQM--0207, FQM--481, FQM--2445 and P06-FQM-01735 (Junta de Andaluc\'{\i}a) is gratefully appreciated. B.O.S. acknowledges the support of Ministerio de Educaci\'on y Ciencia under the program FPU. Support by the Austrian Science Foundation (FWF) through SFB 15 is acknowledged. We thank Andrew Daley, Ernesto Est\'evez-Rams, Markus M\"uller, Johannes Schachenmayer and Peter Zoller for fruitful discussions.
\end{acknowledgments}

\bibliographystyle{apsrev}

\end{document}